\documentstyle[12pt,epsf,epsfig,cite]{article}
\def\Journal#1#2#3#4{{#1} {\bf #2}, #3 (#4)}

\def\NPB{{\em Nucl. Phys.} B}
\def\PLB{{\em Phys. Lett.}  B}
\def\PRL{\em Phys. Rev. Lett.}
\def\PRD{{\em Phys. Rev.} D}

\def\PR{\em Phys. Rep.}

\def\NAT{\em Nature}
\def\APP{\em Astropart. Phys.}

\def\RMP{{\em Rev. Mod. Phys.}}
\newcommand{\newc}{\newcommand}
\newc\eg{{\it {e.g.}}}  \newc\etal{{\it {et al.}}} \newc\ie{{\it i.e.}}
\newc\etc{{\it {etc}}}  

\newc{\mplanck}{M_{\rm P}}      \newc{\mpl}{M_{\rm Pl}}
\newc{\msusy}{M_{\rm SUSY}}      \newc{\ms}{M_{\rm S}}

\newc{\jxf}{J({\xf})}
\newc{\jxfexact}{J_{\rm exact}({\xf})}  \newc{\jxfexp}{J_{\rm exp}({\xf})}
\newc{\VEV}[1]{\langle #1 \rangle}

\newc{\xf}{x_f}
\newc\vrel{v_{\rm rel}}
\newcommand\mchi{m_{\chi}}

\newc\hpm{H^\pm} \newc\hp{H^+} \newc\hm{H^-} 
\newc\sfermion{\tilde f}  \newc\msfermion{m_{\sfermion}}  

\newc\second{{\rm sec}} 

\newc\alphas{\alpha_s}

\newc{\gstar}{g_\ast}           \newc{\gsstar}{g_{s\ast}}
\newc{\geff}{g_{\rm eff}}

\newc{\sthw}{\sin\theta_W}              \newc{\cthw}{\cos\theta_W}
\newc{\bino}{\widetilde B}              \newc{\wino}{\widetilde W_3}
\newc{\higgsinob}{{\widetilde H}^0_b}   \newc{\higgsinot}{{\widetilde H}^0_t}

\newc{\abund}{\Omega h^2}
\newc{\abundchi}{\Omega_\chi h^2}
\newc{\rhocrit}{\rho_{crit}}
\newc{\rhochi}{\rho_{\chi}}

\newcommand\gev{\,\mbox{GeV}}

\newc{\ra}{\rightarrow}
\newc{\beq}{\begin{equation}}
\newc{\eeq}{\end{equation}}
\newc{\bea}{\begin{eqnarray}}
\newc{\eea}{\end{eqnarray}}

\newcommand\lsim{\mathrel{\rlap{\lower4pt\hbox{\hskip1pt$\sim$}}
    \raise1pt\hbox{$<$}}}
\newcommand\gsim{\mathrel{\rlap{\lower4pt\hbox{\hskip1pt$\sim$}}
    \raise1pt\hbox{$>$}}}

\long\def\begincomment#1\endcomment{%
        \begingroup\sf\baselineskip12pt#1\endgroup}

\begin{document}
\begin{titlepage}
\pagestyle{empty}
\baselineskip=21pt
\rightline{CERN--TH/2001-038}
\vskip 0.5in
\begin{center}
{\large {\bf Towards An Accurate Calculation\\ of the Neutralino Relic Density
}}
\end{center}
\begin{center}
\vskip 0.05in
{{\bf Takeshi Nihei}$^1$, 
{\bf Leszek Roszkowski}$^{1,2}$\\
and {\bf Roberto Ruiz de Austri}$^1$}\\
\vskip 0.05in
{\it
$^1${Department of Physics, Lancaster University, Lancaster LA1
4YB, England}\\
$^2${TH Division, CERN, CH-1211 Geneva 23, Switzerland}\\
}
\vskip 0.5in
{\bf Abstract}
\end{center}
\baselineskip=18pt 
\noindent 
We compute the neutralino relic density in the minimal supersymmetric
standard model by using exact expressions for the neutralino
annihilation cross section into all tree-level final states, including
all contributions and interference terms. We find that several final
states may give comparable contributions to the relic density, which
illustrates the importance of performing a complete calculation.  We
compare the exact results with those of the usual expansion method and
demonstrate a sizeable discrepancy (of more than $10\%$) over a
significant range of the neutralino mass of up to several tens of
$\gev$ which is caused by the presence of resonances and new final-state
thresholds.  We perform several related checks and comparisons. In
particular, we find that the often employed approximate iterative
procedure of computing the neutralino freeze-out temperature gives
generally very
accurate results, except when the expansion method is used near
resonances and thresholds.

\vfill
\vskip 0.15in
\leftline{CERN--TH/2001-038}
\leftline{February 2001}
\end{titlepage}
\baselineskip=18pt

%
%
\section{Introduction}\label{intro:sec}

One of the most exciting unresolved issues in particle physics and
cosmology is the nature of dark matter (DM) in the Universe.
Cosmological and astronomical observations have put, over the years,
significant restrictions on its expected properties. In particular, it
is widely expected that a hypothetical candidate should be in the form
of cold DM (CDM), while a combination of several measurements has narrowed
the range of its relic abundance to 
\beq
0.1\lsim \Omega_{\rm CDM}h^2 \lsim 0.3,
\label{eqn:omegah^2}
\eeq
where $\Omega_{\rm CDM}$ is the ratio of the CDM relic density to the
critical density and $h\approx0.7$ is a parameter in the Hubble constant 
$H_0= 100\, h$ km/sec/Mpc~\cite{Hubble}. Constraints from the Big Bang
Nucleosynthesis~\cite{bbn:ref} and, recently, from observations of the 
cosmic microwave background radiation~\cite{Boomerang,Maxima} 
have strongly restricted the
abundance of baryonic DM; for example, Ref.~\cite{TT} quotes the range
$\Omega_b h^2$ $=$ $0.02\pm 0.002$.
Therefore most of the DM in the Universe is expected to be non-baryonic.

From a particle physics point of view, CDM is most naturally made of
some weakly interacting (stable) massive particles (WIMPs).
Supersymmetry (SUSY)~\cite{MSSM}, which over the years has emerged as a leading
candidate for new physics beyond the Standard Model (SM), predicts
that, in the presence of R-parity, the lightest supersymmetric
particle (LSP) is stable. This alone makes the LSP an interesting
candidate for a WIMP and DM in the Universe.

A SUSY candidate for the LSP WIMP is by no means unique. In models coupled
to gravity there exists the gravitino, the SUSY partner of the
graviton. The gravitino has long been known to be a potential
candidate for DM, although it generically suffers from a well-known
``gravitino problem''~\cite{gravitinoprob:ref}.
In SUSY models that incorporate the Peccei--Quinn solution to the
strong CP problem, there exists the axino, the fermionic partner of the
axion. The axino has recently been shown to be an attractive and
well-motivated alternative candidate for the LSP and CDM~\cite{ckkr}. A
number of more speculative possibilities, stemming from, for example,
string theories, 
have also been proposed. However, in some sense the simplest choice
is to consider one of the superpartners of the SM particles as a
potential LSP. Among these, the lightest neutralino $\chi$ stands out
as probably the most natural and attractive candidate for the LSP and
DM~\cite{Goldberg,neutralino-dm}. In this paper we will focus on this case. 

In order to be able to perform a reliable comparison between 
theoretical predictions
and improving measurements of the relic abundance and limits from
underground DM searches, a precise calculation of the relic density
becomes necessary. The  constraint~(\ref{eqn:omegah^2})  is known to
often provide a strong restriction on allowed combinations of
SUSY masses and couplings. For example, in the minimal supergravity
model (also dubbed CMSSM) the condition $\abundchi<0.3$ provides a
strong upper bound of a few hundred~$\gev$ 
on the masses of gauginos and scalars over an overwhelming fraction of
the parameter space~\cite{rr93,kkrw}. When combined with recent LEP
and other data, the remaining  CMSSM parameter space becomes
highly restricted~\cite{efgos01}.
In the future, the range~(\ref{eqn:omegah^2}) is
expected to be significantly narrowed by MAP and Planck,
which in turn will lead to even stronger constraints on allowed
combinations of SUSY masses and couplings.

One of the key steps in computing the relic density of the
neutralino involves calculating the thermal average of its
annihilation cross section times the relative velocity. This calculation
is technically rather involved and, therefore, in most of the
literature, approximate methods have been used. These methods are
based on expanding the thermal average in powers of $x\equiv T/\mchi$,
where $T$ is the temperature of the thermal bath and $\mchi$ is the
neutralino mass. One usually computes only the first two terms of the
expansion.  It has been well known that the expansion fails badly near
$s$-channel resonances~\cite{gs91,gg91,an93,lny93} and near new
final-state thresholds~\cite{gs91,gg91}. In particular, it was
emphasized in Ref.~\cite{an93} that, owing to the very narrow width of
the lightest SUSY Higgs boson $h$, the
error can be as large as a few orders of magnitude very close to the
$h$-resonance. However, even
though the usual expansion is generally expected to be accurate enough
further away from resonances and thresholds, no detailed study of this
point has yet been done. 

On the other hand, while a general formalism for computing the thermal
average precisely was derived a long time ago~\cite{gg91,swo88}, a full
set of exact, general and explicit analytic expressions for the
neutralino pair-annihilation cross section is still not available in
the literature. (The thermal average can be obtained by performing a
single integral over the cross section, as we will see below.)  The
cross sections for the neutralino pair-annihilation into the SM
fermion-pair ($f\bar f$) and the lightest Higgs--boson pair ($hh$)
final states were given in Ref.~\cite{lny93}, and for the $WW$ and
$ZZ$ final states in Ref.~\cite{gkt90}.  The annihilation into $f\bar
f$ is often most important.
However, other final
states can also play a considerable role, depending on the model. 

We have derived a full set of exact analytic expressions for the cross
section of the neutralino pair-annihilation in the minimal
supersymmetric standard model (MSSM). We have made no simplifying
assumptions about the degeneracy of the left- and right--sfermion
masses, and we included all tree-level final states and all intermediate
states. We have kept finite widths in $s$-channel resonances. While
interference terms and usually sub-dominant channels are sometimes
neglected in the literature, in some cases they may give significant
contributions.  In particular, the channels involving gauge--Higgs
boson may in some cases be sizeable or even dominant. One example
where this is the case is the $W^\pm H^\mp$ final state, and in this
paper we give expressions for the cross section for the chargino exchange
contribution.  A full set of all the expressions for all the final
states will be presented elsewhere~\cite{nrr2}.

As a check, we have performed a numerical comparison of our 
cross section with the results obtained by using the recently released code
DarkSusy~\cite{darksusy00}. We have found, for the same values of input
parameters, an impressive agreement, at the level of a few per cent, for
all the annihilation channels, which we find reassuring.

We have also made a comparison of the relic abundance computed using
our exact formulae with the one obtained using our expansion formulae.
So far, no such detailed analysis has been
presented in the literature in the case of the MSSM. In Ref.~\cite{lny93} an analogous
comparison was made in the context of the highly constrained minimal
supergravity scheme and only relatively close to resonances.  We
confirm that the expansion gives highly inaccurate results near
resonances and new thresholds. We also show that very far from such
cases the error is typically rather small, of the order of a few per cent.
However, we find that, 
because of the existence of several resonances ($Z$ and the
Higgs bosons), the expansion produces large errors, compared to an exact
treatment, over a sizeable range of $\mchi$,  even of a several tens
of~$\gev$. We therefore conclude that the widely used method of
expansion may lead to significant errors in a sizeable fraction of the
neutralino mass which is typically expected not to significantly
exceed $\sim200\gev$ by various naturalness criteria~\cite{chiasdm,dg}.

In the current version of our code we have used a popular approximate
iterative method of determining the freeze-out point. We have examined
the accuracy of the procedure and compared our results with the ones
obtained using
DarkSusy. We have found that the method in general works very well
when both exact and expansion methods are used, except in the latter
case near resonances and thresholds where it may even break down. The
iterative procedure can be safely applied in such special cases,
provided exact expressions for the cross section are used and much
care is paid to properly integrating the thermal average numerically.

%
\section{Calculation of the Relic Density}\label{relicdensity:sec}

The  relic abundance of some stable species $\chi$ is given by 
$\Omega_{\chi} \equiv \rho_\chi/{\rho_{crit}}$,
where $\rho_\chi$ is the relic's (mass)
density and $\rho_{crit}\equiv 3 H_0^2/8\pi G = 1.9\times 10^{-29}\,
(h^2)\,g/cm^3$ is the critical density. It can be computed by solving 
the Boltzmann
equation, which describes 
the time evolution of the co-moving number density $n_\chi$ 
in the expanding Universe,\footnote{
For a review of calculations
of the relic density, see, \eg, Refs.~\cite{kt90,jkg96}.}
\begin{eqnarray}
\frac{d n_\chi}{dt} & = & - 3 H n_\chi
- \langle\sigma v_{\rm M\o l}\rangle 
\left(n_\chi^2 - (n_\chi^{\rm eq})^2\right), 
\label{eqn:Boltzmann-eq}
\end{eqnarray}
where $n_\chi^{\rm eq}$ is the number density that the species would 
have in thermal equilibrium, 
$H(T)$ is the Hubble expansion rate,  
$\sigma(\chi \chi
\rightarrow {\rm all})$ denotes the cross section of the species
annihilation into ordinary particles, 
$v_{\rm M\o l}$ is a so-called M{\o}ller velocity~\cite{gg91} which is 
the relative velocity
of the annihilating particles, and
$\langle\sigma v_{\rm M\o l}\rangle$ represents the thermal average of 
$\sigma v_{\rm M\o l}$ which will be given below. 
In the early Universe, the species $\chi$ were initially in thermal
equilibrium, $n_\chi$ $=n_\chi^{\rm eq}$. When their typical
interaction rate $\Gamma_\chi$ became less than the Hubble parameter, 
$\Gamma_\chi\lsim H$, the annihilation
process froze out. Since then their number in a co-moving volume
has remained basically constant.

In order to calculate the relic density precisely enough, one has to
consider the following ingredients: (i) compute all the contributions
to the annihilation cross section, (ii) use an exact formula for the
thermal average, (iii) rigorously solve the Boltzmann equation, and
(iv) include co-annihilation. (This becomes important when the mass
of the next LSP is nearly degenerate with the LSP
mass~\cite{gs91,Mizuta-Yamaguchi,ino-coan,stau-coan}.)  In this work,
we will concentrate on points (i) and (ii) but will also examine to
some extent point (iii). In particular, we will compare the usually
used expansion formulae with the exact ones. We will not consider here
the effect of co-annihilation.

An often used, approximate, although in general quite accurate (see later),
solution to the Boltzmann equation 
is based on solving iteratively the  equation
\begin{eqnarray}
x_f^{-1} & = & \ln \left( \frac{m_\chi}{2 \pi^3} \sqrt{\frac{45}{2g_* G_N}}
\langle\sigma v_{\rm M\o l}\rangle({x_f})\, x_f^{1/2} \right),
\label{eqn:freeze-out-temperature}
\end{eqnarray}
where $g_*$ represents the effective
number of degrees of freedom at freeze-out ($\sqrt{g_*}\simeq 9$) 
and $G_N$ is the Newton constant.
Typically one finds that the freeze-out point
$x_f\equiv T_f/\mchi$ is roughly given by $1/20$. 
The relic density $\rho_\chi=m_\chi n_\chi$
at present is given by
\begin{eqnarray}
\rho_\chi & = & \frac{1.66}{M_{\rm Pl}} 
\left(\frac{T_\chi}{T_\gamma}\right)^3 T_\gamma^3 \sqrt{g_*}
\frac{1}{ \jxf},
\label{eqn:relic-density}
\end{eqnarray}
where $M_{\rm Pl}$ denotes the Planck mass, $T_\chi$ and $T_\gamma$
are the present temperatures of the neutralino and the photon,
respectively. The suppression factor $(T_\chi/T_\gamma)^3$ $\approx$
$1/20$ follows from entropy conservation in a comoving
volume~\cite{reheating_factor}. Finally, $\jxf$ is given by
\beq
\jxf\equiv \int_0^{x_f}dx \langle\sigma v_{\rm M\o l}\rangle(x),
\label{jxfdef:eq}
\eeq
where 
$x=T/m_\chi$, as defined earlier.

Below we will concentrate on computing $\jxf$ and in particular on
the thermally averaged
annihilation cross section times velocity $\langle \sigma v_{\rm M\o l}
\rangle$. 
A proper definition which
includes separate thermal baths for both annihilating particles
is given by~\cite{swo88,gg91}
\begin{equation}
\langle \sigma v_{\rm M\o l} \rangle(T)=
\frac{\int d^3p_1 d^3p_2\, \sigma v_{\rm M\o l}\, e^{-E_1/T} e^{-E_2/T}}
{\int d^3p_1 d^3p_2\,  e^{-E_1/T} e^{-E_2/T} }
\label{eq:sigmavdef}
\end{equation}
where $p_1= (E_1, {\bf p}_1)$ and  $p_2= (E_2, {\bf p}_2)$ 
are the 4-momenta of the two colliding particles, and $T$ 
is the
temperature of the bath.\footnote{
Note that in many cases one uses another definition of
$<\sigma v_{\rm M\o l}>$ which involves a single thermal bath for both
neutralinos. Compare, \eg, Refs.~\cite{kt90,dn93,jkg96}.
}
The above expression can be reduced to a one-dimensional
integral which can be written 
in a Lorentz-invariant form as~\cite{gg91}
\begin{eqnarray}
\langle\sigma v_{\rm M\o l}\rangle(T) & = & 
\frac{1}{8 m_\chi^4 T K_2^2(m_\chi/T)} 
\int_{4 m_\chi^2}^\infty ds \, \sigma(s) (s-4m_\chi^2)\sqrt{s}
K_1\left(\frac{\sqrt{s}}{T}\right),
\label{eqn:thermal-average}
\end{eqnarray}
where  $s=(p_1+p_2)^2$ is a usual Mandelstam variable and
$K_i$ denotes the modified Bessel function of order $i$.

In computing the cross section
$\sigma(s)$, it is convenient to introduce a Lorentz-invariant
function $w(s)$~\cite{swo88} 
\beq w(s)= \frac{1}{4} \int d\, {\rm LIPS}\, |{\cal A} (\chi \chi
\rightarrow {\rm all})|^2
\label{wdef:eq}
\eeq
where $|{\cal A} (\chi \chi \rightarrow {\rm all})|^2$ 
denotes the absolute square of the reduced matrix
element for the annihilation of two $\chi$ particles, averaged over
initial spins and summed over final spins. The function $w(s)$ is
related to the annihilation cross section via~\cite{lny93}
\beq
w(s)=  \frac{1}{2} \sqrt{ s (s-4\mchi^2)}\, \sigma(s).
\label{wtosigma:eq}
\eeq

For a general annihilation process into a two-body final state
$\chi\chi\rightarrow f_1 f_2$, the function $w(s)$
can be written as
\begin{eqnarray}
w(s)&=&\frac{1}{32\,\pi}\sum \bigg[
\theta\left(s-(m_{f_1}+m_{f_2})^2 \right)\,
\beta_f(s,m_{f_1},m_{f_2})\,\tilde{w}_{f_1f_2}(s)\bigg], 
\label{wtowtilde:eq}
\end{eqnarray}
where the summation goes over all possible channels, and 
\begin{eqnarray}
\tilde{w}_{f_1f_2}(s)\equiv \frac{1}{8\pi}\int \!d\Omega \,|{\cal A}(\chi \chi
\rightarrow f_1 f_2)|^2, 
\label{wtildedef:eq}
\end{eqnarray}
where
\begin{eqnarray} 
\beta_f(s,m_{f_1},m_{f_2})\equiv
\left[1-\frac{(m_{f_1}+m_{f_2})^2}{s}\right]^{1/2}
\left[1-\frac{(m_{f_1}-m_{f_2})^2}{s}\right]^{1/2}. 
\label{kdef:eq}
\end{eqnarray}

Computation of $w(s)$ is in general rather involved and final analytic
expressions exhibit considerable complexity, as will be illustrated in
the next Section. This is one reason why in most of the
literature one uses the well-known approximation in terms of 
expansion in powers of
$x$,
$\langle\sigma v_{\rm M\o l}\rangle \simeq a + bx$.
Using the definition~(\ref{eq:sigmavdef}), the expansion
reads~\cite{swo88}
\begin{equation}
<\sigma v_{\rm M\o l}>=\frac{1}{m_\chi^2}
\left[w-\frac{3}{2}\left(2w-w'\right) x + 
{\mathcal O} (x^2)\right]_{s=4m_\chi^2} \equiv a + bx + {\mathcal
O}(x^2)
\label{expansion:eq}
\end{equation}
where $w^\prime(s)$ denotes $d\,w(s)/d\,(s/4\mchi^2)$. 

Note that in case of the expansion of $<\sigma v_{\rm M\o l}>$ defined
in terms of a single thermal bath for both neutralinos, the
corresponding coefficient $a^\prime$ is the same as
Eq.~(\ref{expansion:eq}) while $b^\prime= b+
\frac{3}{2}a$~\cite{srednicki:ref}.  For more details, see, \eg,
Ref.~\cite{simple:ref}.

The expansion method is widely used not only because it is
computationally somewhat less
involved but also because it is expected to be relatively
accurate. This can seen by 
examining the behavior of 
the integrand in Eq.~(\ref{eqn:thermal-average}). For
a massive particle like the neutralino for which 
$T\lsim m_\chi/20$ and $\sqrt{s}\geq 2 m_\chi$, 
the argument of the function $K_1$ is much larger than
unity. Since $K_1(y) \sim \sqrt{\pi/2y} e^{-y}$ 
at $y\gg 1$,  
the thermal average~(\ref{eqn:thermal-average}) 
can be written as a convolution of the cross
section with a function $f(s)$:
\begin{eqnarray}
\langle\sigma v_{\rm M\o l}\rangle & \approx & 
\int_{4 m_\chi^2}^\infty ds \, \sigma(s) f(s),
\label{eqn:convolution}
\end{eqnarray}
where $f(s)$ has an exponential suppression factor, 
$f(s) \propto e^{-\sqrt{s}/T}$.
Thus one would expect that the usual 
expansion in powers of $x$
should converge quickly~\cite{lny93}.

One should however use this argument with some care. First, the
annihilation cross section may change rapidly with $s$. It is
well-known that this happens, \eg, near $s$-channel resonances and
thresholds of new final states. In such cases the expansion method
produces large errors~\cite{gg91,gs91,an93,lny93}. Far away from 
such singular points the expansion is expected to give accurate
results but, to our knowledge, this has never been explicitly verified
in the literature. Finally, one would want to examine actually how far from
resonances one can start relying on the usual approximation. As we
will see later, the range of $\mchi$ where the expansion fails to be
reasonably accurate can be actually quite large.

%
\section{Analytic Results}\label{analytic:sec}

As stated in the Introduction, we have derived a full set of exact,
analytic expressions for the cross sections for the neutralino
pair-annihilation processes into all allowed (tree-level) final states
in the general MSSM. We have included all contributing diagrams as
well as all interference terms and kept finite widths of all
$s$-channel resonances. We have made no simplifying assumptions about
sfermion masses although we assumed that there are no CP violating
phases in SUSY parameters. Here we will give one analytic example of
our exact expressions. A full set of results for all the final states
will be given elsewhere~\cite{nrr2}.

Next, starting from the exact expressions we have computed the
coefficients $a$ and $b$ using Eq.~(\ref{expansion:eq}) for all the
partial annihilation channels.  In the literature there exist several
analytic formulae for the expansion coefficients,
including~\cite{erl90,os91,dn93,jkg96}, but, due to different
conventions and complexity of expressions, comparison is not always
doable. We have checked our results for the $a$-coefficients in
appropriate limits against published results and agreed in all cases.
In the next Section we will present a numerical comparison of the
approximation in terms of the usual expansion with our exact results.

Let us begin by introducing the relevant MSSM parameters.\footnote{We follow
the convention of Ref.~\cite{gh}.}
The lightest neutralino is a mass eigenstate given by a linear 
combination of two neutral gauginos and two neutral higgsinos
\begin{eqnarray}
\chi & = & N_{11} \tilde{B} + N_{12} \tilde{W}_3 
+ N_{13} \tilde{H}^0_1 + N_{14} \tilde{H}^0_2. 
\label{eqn:chi-1}
\end{eqnarray}
The neutralino mass matrix is determined by 
the $U(1)_Y$ and $SU(2)_L$ gaugino mass parameters $M_1$
and $M_2$ (and we impose the usual GUT relation 
$M_1=\frac{5}{3}\tan^2\theta_W  M_2 $),  
the Higgs/higgsino mass parameter $\mu$ and 
$\tan \beta=v_2/v_1$ which is the ratio of the vacuum expectation values of 
the two neutral Higgs fields. 

There are two neutral scalar Higgs bosons $h$ and $H$ and one
pseudoscalar $A$ plus a pair of charged Higgs $H^\pm$. (We will
typically suppress the Higgs charge assignment except 
where this may lead to ambiguities.)  We use
expressions given in Ref.~\cite{rc-higgsmass}
for computing radiatively corrected Higgs masses.

Other relevant parameters which determine the masses of scalars and
various couplings are the squark soft mass parameters $m_Q$, $m_U$ and
$m_D$, the slepton soft mass parameters $m_L$ and $m_E$, and the
pseudoscalar mass $m_A$. We also include the trilinear terms $A_i$
($i=t,b,\tau$) of the third generation which are important in
determining the masses and couplings for the stop, sbottom and stau sfermions,
respectively.

There are a number of final states into which the neutralino can
pair-annihilate. They include: $\chi \chi\rightarrow f \bar{f}$, $WW$,
$ZZ$, $Zh$, $ZH$, $ZA$, $W^\pm H^\mp$, $AA$, $AH$, $Ah$, $HH$, $Hh$,
$hh$ and $H^+ H^-$.  Among these, in the gaugino region, fermion pair
$f \bar{f}$ final states usually give dominant contributions, unless
the exchanged sfermion masses $\msfermion$ are large and in some
special cases, as discussed earlier. These final states are also
always kinematically allowed (except for $t \bar t$) but, due to the $s$-wave
suppression~\cite{Goldberg}, their cross section is proportional to
$m_f^2/m_W^2$. 
For the higgsino-like neutralinos
the states $ZZ$ and $WW$ become very important once kinematically
allowed~\cite{gkt90}.  However, these final states can give significant
contributions even in the case of the gaugino-like
neutralino. 
This is because $WW$ and $ZZ$ are not $s$-wave suppressed, unlike the
channels $hh$, $Hh$, $HH$, $AA$, $H^+ H^-$ and $ZA$~\cite{jkg96}. What
is interesting is that the cross sections for the other $s$-wave
unsuppressed states $W^\pm H^\mp$, $Zh$, $ZH$, $Ah$ and $AH$, once
kinematically allowed, can be even larger than those of $WW$ and
$ZZ$. In fact, they can even dominate over those of $f\bar{f}$.  These
points will be illustrated with numerical examples in
Section~\ref{numerical:sec}.

Here, we present a set of expressions for a $t$- and $u$-channel
chargino ($\chi^\pm_{1,2}$) exchange contribution to the process
$\chi\chi$ $\rightarrow$ $W^+ H^-$.  This contribution is often
dominant. (There are also $s$-channel diagrams involving $h$, $H$ and
$A$ exchange. Since typically $m_h\ll m_H\simeq m_A \simeq m_{H^\pm}$,
the Higgs resonances will be  outside of the kinematically allowed region.)
The
function $\tilde{w}$, Eq.~(\ref{wtildedef:eq}) in this case reads 
\begin{eqnarray}
\tilde{w}_{W^{+}H^{-}}^{\chi^\pm} & = &
\frac{1}{m_{W}^{2}}\sum_{i,j=1}^{2}
\bigg[ m_{\chi_i^\pm}\,m_{\chi_j^\pm} I^{(2)}_{ij}
      + m_{\chi} m_{\chi_i^\pm} I^{(1)}_{ij} + I^{(0)}_{ij} \bigg],
\label{eqn:w_WH}
\end{eqnarray}
where
\begin{eqnarray}
\!\!\!\! I^{(2)}_{ij} & = & 
(C_{+i}^{HW^*}C_{+j}^{HW}+D_{+i}^{HW^*}D_{+j}^{HW}) \nonumber \\
 & & \hspace{5mm} \times \Big[ - {\mathcal T}_2 
       - (s-m_W^2-m_{H^{\pm}}^2-2\,m_{\chi}^2)\,{\mathcal T}_1
       + G_{WH}^{T(1)}\,{\mathcal T}_0 
       + 6\,m_W^2 m_{\chi}^2\,{\mathcal Y}_0 \,\Big] \nonumber \\
 & & + \ (C_{+i}^{HW^*}C_{+j}^{HW}-D_{+i}^{HW^*}D_{+j}^{HW})
 \Big[ \,6\,m_W^2 m_{\chi}^2\,{\mathcal T}_0
       - {\mathcal Y}_2 + G_{WH}^{Y(1)}\,{\mathcal Y}_0 \,\Big], 
\label{eqn:I_2} \\
\!\!\!\! I^{(1)}_{ij} & = & 
{\rm Re}(C_{+i}^{HW^*}C_{-j}^{HW}+D_{+i}^{HW^*}D_{-j}^{HW})
 \Big[ -2\,{\mathcal T}_2 
       + 2\,(2\,m_{\chi}^2-m_W^2)\,{\mathcal T}_1 \nonumber \\
 & & \hspace{2.3cm} - \ 2(m_{\chi}^2-m_W^2)(m_{\chi}^2+2\,m_W^2){\mathcal T}_0
                    + 2\,m_W^2\,{\mathcal Y}_1 
                    - \ G_{WH}^{Y(2)}\,{\mathcal Y}_0 \,\Big] \nonumber \\
 & & + \ {\rm Re}(C_{+i}^{HW^*}C_{-j}^{HW}-D_{+i}^{HW^*}D_{-j}^{HW})
 \Big[ \,6\,m_W^2\,{\mathcal T}_1
       + 6\,m_W^2(m_{\chi}^2-m_{H^{\pm}}^2){\mathcal T}_0 \nonumber \\
 & & \hspace{4.5cm} - \ 2\,{\mathcal Y}_2 
                    - (s-m_{H^{\pm}}^2-4\,m_W^2){\mathcal Y}_1
                    - G_{WH}^{Y(3)}\,{\mathcal Y}_0 \,\Big], 
\label{eqn:I_1} \\
\!\!\!\! I^{(0)}_{ij} & = & 
(C_{-i}^{HW^*}C_{-j}^{HW}+D_{-i}^{HW^*}D_{-j}^{HW})
 \Big[ \,(s-m_{\chi}^2-2\,m_W^2)\,{\mathcal T}_2
       - G_{WH}^{T(2)}\,{\mathcal T}_1  \nonumber \\
 & &   \hspace{5.25cm} - \ G_{WH}^{T(3)}\,{\mathcal T}_0 
       + (s-2\,m_W^2)\,{\mathcal Y}_2
       + G_{WH}^{Y(4)}\,{\mathcal Y}_0 \,\Big] \nonumber \\
 & & + \ (C_{-i}^{HW^*}C_{-j}^{HW}-D_{-i}^{HW^*}D_{-j}^{HW})
 \Big[ \,6 \,m_W^2 m_{\chi}^2\,{\mathcal T}_1
       - m_{\chi}^2\,{\mathcal Y}_2 
       + G_{WH}^{Y(5)}\,{\mathcal Y}_0 \,\Big]. 
\label{eqn:I_0}
\end{eqnarray}

Obviously the total contribution to $w(s)$ from $W^\pm H^\mp$ is twice that
from $W^+ H^-$.
The symbols $C_{\pm i}^{HW}$ and $D_{\pm i}^{HW}$ in the above equations
are combinations of several coupling constants:
\begin{eqnarray}
C_{\pm i}^{HW} & \equiv& C_{S}^{\chi_{i}^{+}\chi H^{-}}
     C_{V}^{\chi_{i}^{+}\chi W^{-^{*}}}\pm \:C_{P}^{\chi_{i}^{+}\chi
     H^{-}} C_{A}^{\chi_{i}^{+}\chi W^{-^{*}}}, \\ 
D_{\pm i}^{HW} & \equiv& C_{S}^{\chi_{i}^{+}\chi H^{-}} C_{A}^{\chi_{i}^{+}\chi
     W^{-^{*}}}\pm \:C_{P}^{\chi_{i}^{+}\chi H^{-}}
     C_{V}^{\chi_{i}^{+}\chi W^{-^{*}}}.
\label{eqn:Cpm_and_Dpm}
\end{eqnarray}
The relevant couplings describe neutralino-chargino-$W^-$/$H^-$
interaction terms in the Lagrangian as
\begin{eqnarray}
     {\cal L}_{\rm int} &=& \sum_{i=1}^{2}
    \overline{\chi^-_i} \gamma^\mu
    \left( C_V^{\chi_i^+ \chi W^-}
           -C_A^{\chi_i^+ \chi W^-}\gamma_5 \right)
     \chi W_\mu^- \nonumber \\
    & & + \, \sum_{i=1}^{2}
    \overline{\chi^-_i}
    \left( C_S^{\chi_i^+ \chi H^-}
           -C_P^{\chi_i^+ \chi H^-}\gamma_5 \right)
     \chi H^- + \,{\rm h.c.}
\label{eqn:couplings}
   \end{eqnarray}
in the convention of Ref.~\cite{gh}. 
The two pairs of charginos are denoted by 
$\chi^\pm_i$ ($i= 1,2$) and $m_{\chi_i^\pm}$ represent their masses. 
The functions ${\mathcal T}_k$ $=$ ${\mathcal T}_k$($s$, $\!
m_{\chi}^2$, $\! m_{H^{\pm}}^2$, $\! m_W^2$, $\! m_{\chi_i^\pm}^2$, $\!
m_{\chi_j^\pm}^2$), ${\mathcal Y}_k$ $=$ ${\mathcal Y}_k$($s$, $\!
m_{\chi}^2$, $\! m_{H^{\pm}}^2$, $\! m_W^2$, $\! m_{\chi_i^\pm}^2$, $\!
m_{\chi_j^\pm}^2$), $k$ $=$ $0,1,2$, and the functions $G_{WH}^{T(1-3)}$,
$G_{WH}^{Y(1-5)}$ are listed in the Appendix.

%
\section{Numerical Results}\label{numerical:sec}

\begin{figure}[t]
\hspace*{-0.5cm}
\begin{minipage}{8in}
\epsfig{file=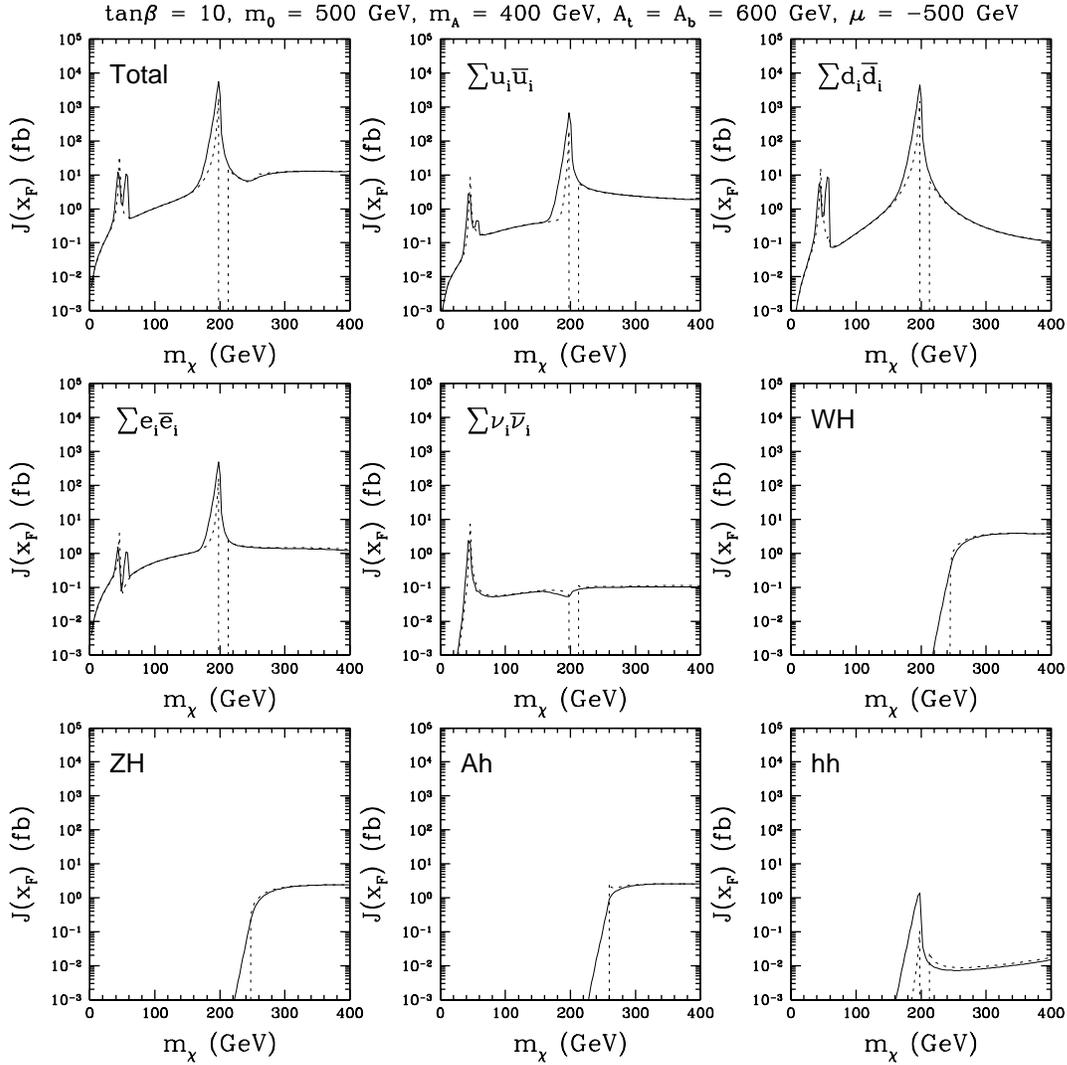,height= 6in}
\end{minipage}
\caption[fig1]{
The total value of $\jxf$, Eq.~(\protect{\ref{jxfdef:eq}}), and several partial
contribution are shown in separate windows as a function of $m_\chi$
for $\tan \beta=10$, $m_0\equiv m_Q=m_U=m_D=m_L=m_E=500\gev$, $m_A=400\gev$,
$A_t=A_b=600\gev$ and $\mu=-500\gev$. The solid lines represent
the exact results, while the dotted ones correspond to the
expansion~(\protect{\ref{expansion:eq}}. 
Notice that the final states $W^\pm H^\mp$, $ZH^0$ and $A
h$, once kinematically allowed,
give comparable contributions to the $f\bar{f}$ channels.}
\label{fig:nrr-fig1} 
\end{figure}

\begin{figure}[t]
\hspace*{-0.5cm}
\begin{minipage}{8in}
\epsfig{file=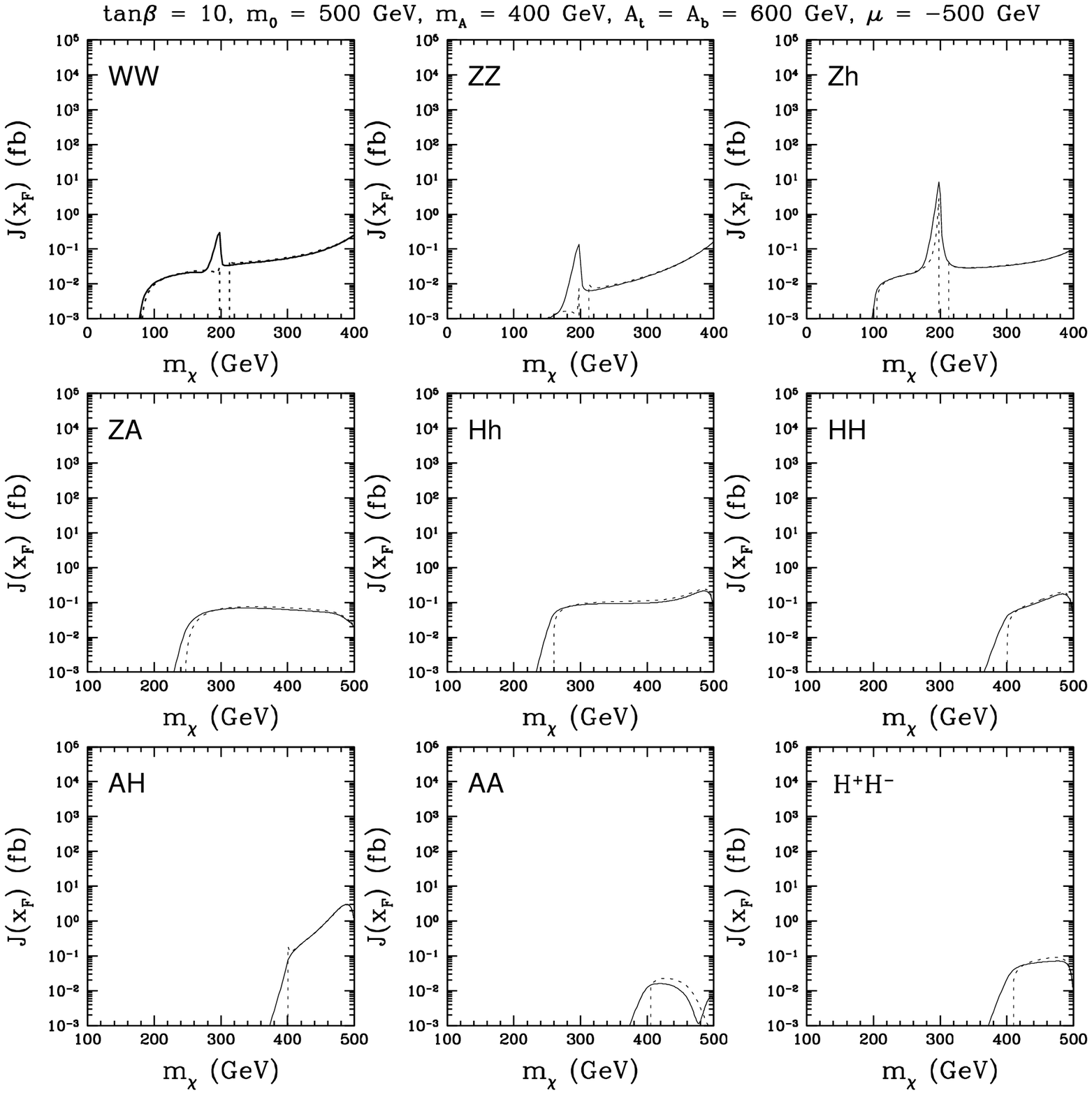,height= 6in}
\end{minipage}
\caption[fig2]{
The same as in Fig.~\protect{\ref{fig:nrr-fig1}} but for mostly subdominant
channels. Notice that in the lower two rows the horizontal axis has
been shifted by 100 GeV.
}
\label{fig:nrr-fig2} 
\end{figure}

In this Section we will illustrate the points made above by presenting
several numerical results. We will concentrate on the gaugino-like
neutralino. We remind the reader that a nearly pure gaugino (bino) is
the most 
natural choice for the CDM both in the MSSM~\cite{chiasdm} and in the CMSSM
where in most cases the LSP comes out to be a
nearly pure bino~\cite{an93,rr93,kkrw}. 

We begin by plotting in Figs.~\ref{fig:nrr-fig1}
and~\ref{fig:nrr-fig2} the function $\jxf$, Eq.~(\ref{jxfdef:eq}),
computed using exact and expansion methods as a function of
$m_\chi$. We take $\tan \beta=10$, $m_0\equiv
m_Q=m_U=m_D=m_L=m_E=500\gev$, $m_A=400\gev$, $A_t=A_b=600\gev$ and
$\mu=-500\gev$. To obtain sparticle and Higgs mass spectra we have
used the code DarkSusy.  While several analytic expressions and
numerical codes (\eg, SUSPECT~\cite{ref:suspect}) are available in the
literature, here we have used DarkSusy to facilitate the comparison of
our results for the cross sections with those produced by the package.
For the values given above we obtain the following Higgs masses: $m_h=
117.3\gev$, $m_H=400.5\gev$ and $m_{H^\pm}= 407.9\gev$.

We show both the total and the
individual contributions from all the final states.
The solid and dotted lines represent the exact  and the
expansion results, respectively. For the sake of comparison of both
methods, for now we do not
impose experimental constraints. Note that $\mchi$ increases with
increasing $M_2$ while we keep $\mu$ fixed at a rather large value.
This means that in these Figures (and also the following ones below),
the neutralino is mostly a gaugino. Even at $\mchi=400\gev$,
$N^2_{11}+N^2_{12}>0.9$. (However, between $400\gev$ and $500\gev$ the
LSP turns into a nearly pure 
higgsino.) In this region the effect of co-annihilation
with the lightest chargino and the next lightest neutralino, which
we do not consider here, is not important and can be safely neglected.

For each final state, we compute $\jxf$, Eq.~(\ref{jxfdef:eq}),
numerically via Eq.~(\ref{eqn:thermal-average}) by using our exact
expressions for the cross sections ($\jxfexact$). Next, we use the
cross sections to compute the first two expansion coefficients $a$ and
$b$, as in Eq.~(\ref{expansion:eq}), and to compute $\jxfexp= a x_f +
\frac{1}{2} b x_f^2$. For consistency, in both cases we determine the
freeze-out point $\xf$ by employing the iterative
procedure~(\ref{eqn:freeze-out-temperature}). We will come back to the
accuracy of the iterative procedure below.

In the behavior of the total $\jxf$ one can clearly recognize three
peaks which are due to the $s$-channel $Z$, $h$ and $A/H$ exchange
contributions.  In particular, we can see that the $A$-resonance is
very broad. (Due to a near mass degeneracy $m_H\simeq m_A$, the
usually narrower $H$-pole is `burried'' underneath the $A$-pole.)
This is caused by the amplification of the coupling
$Ab\bar{b}\sim\tan\beta$ at larger $\tan\beta$ and the fact that, in
the resonance region, the amplitude-square for the $A$-exchange is
proportional to $s$, unlike the case of
$h$ and $H$ where it goes like $s-4m_\chi^2\approx 0$ at
$\sqrt{s}\approx 2m_\chi$.  The $Z$-pole is also clearly visible in
all the partial $f\bar{f}$ final states while the Higgs poles do not
contribute to the $\nu\bar\nu$ final state since neutrinos do not
couple to Higgs bosons. The $H$-pole also gives a visible enhancement
to the $WW$, $ZZ$ and $hh$ final states but not to $W^\pm H^\mp$,
$ZA$, $HH$, $Hh$, $AA$ and $H^+ H^-$ which do not become kinematically
allowed until $\mchi > m_H/2$.  Likewise, the $A$-pole shows a
resonance in the $Zh$ final state but not in $W^\pm H^\mp$, $ZH$, $AH$
and $Ah$.

For the pseudoscalar Higgs exchange in $\chi\chi\rightarrow f\bar{f}$,
the expansion method gives a discontinuity very close to the $A$-pole,
$m_\chi\approx m_A/2=200\gev$. For this channel, which is clearly
dominant in the resonance region, the $b$-coefficient is large and
negative and the expansion method and iterarative procedure break
down.  This can be clearly seen in all the $f\bar f$ windows
(including $\nu\nu$) and also in the total value of $\jxfexp$.  In the
close vicinity of the poles, one can see the well-known large
difference~\cite{gs91,an93,lny93} (note a log scale) between
$\jxfexact$ and $\jxfexp$ while away from the poles, both
contributions are similar.

We can also notice that $\jxfexact$ and $\jxfexp$ show a significantly
different enhancement in the total values and in
the up-type quark-pair final state in the region
$m_\chi\gsim m_t=175\gev$ where the $t\bar{t}$ final state
becomes kinematically allowed. In particular, $\jxfexact$ increases
gradually starting from $\mchi$ some $10\gev$ below
$m_t$ while $\jxfexp$ exhibits a sharp increase only at
$\mchi=m_t$. This is because the exact treatment takes into account the
thermal distribution of momenta of the annihilating neutralinos, while
the expansion method neglects it~\cite{gs91}. The same effect is
visible near the thresholds of all the final states which only become
open at some non-zero value of $\mchi$, especially of those involving one
or two Higgs bosons.

As we stated in Section~\ref{analytic:sec}, once kinematically
allowed, the contributions from the $W^\pm H^\mp$ final state can be
comparable, or even larger (if sfermion masses are increased) than the
$f \bar{f}$ final state. In fact, the $ZH$, $Ah$ and, at much larger
$\mchi$ also $AH$, final state
contributions are not much smaller.  
We can also see that the channels $ZZ$ and $WW$, as well as $Zh$,
despite not being $s$-wave suppressed, are
usually subdominant, except near $\mchi\simeq m_A/2$ because of the
pseudoscalar Higgs exchange contributing to those channels, and when
the higgsino admixture of the neutralino increases which is the case
with increasing $\mchi$ while keeping $\mu$ fixed. 

As a way of verifying our results, we have  performed a numerical
comparison of our exact results for the cross section (more precisely
of the function $w(s)$) with the same quantity computed using
DarkSusy.  We have found an excellent agreement at the level of a few
per cent for all the tree-level final states when we fixed the same
values of input parameters.
This gives us some confidence that the (highly complex) expressions
that we have derived (and also those in DarkSusy), are correct.

%
\begin{figure}[t]
\vspace*{-4cm}
\hspace*{0.5cm}
\unitlength 1mm
\epsfxsize=15.0cm
\begin{minipage}{8in}
\epsfig{file=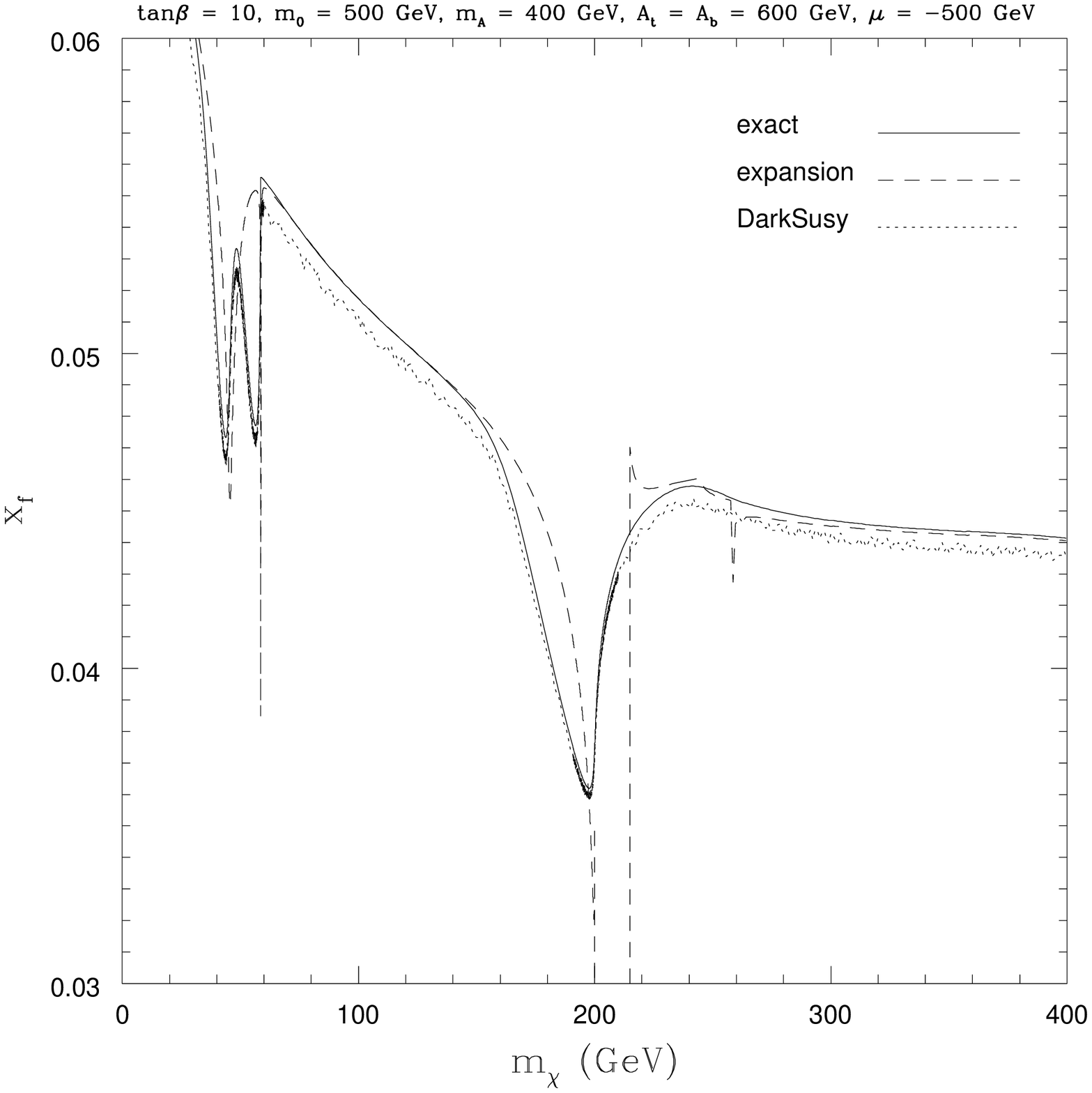,height= 6in}
\end{minipage}
\caption[fig3]{ 
The freeze-out point $x_f$ as a function of $\mchi$. The solid
and dashed lines corresponds to the iterative
procedure~(\protect{\ref{eqn:freeze-out-temperature}}) with $\langle
\sigma v_{\rm M\o l} \rangle$ computed exactly
(\protect{\ref{eq:sigmavdef}}) and in terms of the expansion
(\protect{\ref{expansion:eq}}), respectively. 
For comparison, the dotted line has
been obtained using DarkSusy.
}
\label{fig:xf}
\end{figure}

As mentioned above, in the current version of our numerical code we
have used the iterative method~(\ref{eqn:freeze-out-temperature})
which is widely used in the literature. We applied it to both the
exact formula for $\langle \sigma v_{\rm M\o l} \rangle$,
Eq.~(\ref{eqn:thermal-average}), and to the usual
approximation~(\ref{expansion:eq}). The results are presented in
Fig.~\ref{fig:xf} as a solid and dashed line, respectively. For
comparison, we also display $\xf$ using DarkSusy (dotted line) where
a different numerical procedure is used.  It is clear that there is a
general agreement among the three different procedures and also that
$\xf$ can be only roughly approximated by the usually quoted value
$0.05$. Away from the resonances and thresholds the expansion method
can be safely used in determining $\xf$ while near such special points
this is not the case. In particular, near the $A$-pole the expansion
gives $\langle \sigma v_{\rm M\o l} \rangle<0$ and the iterative
procedure  breaks down. In contrast, the exact numerical
integration~(\ref{eq:sigmavdef}) provides reliable results both near
and away from special points. However, we have carefully verified that this is
the case only when the integration in Eq.~(\ref{eq:sigmavdef}) is performed
with high enough numerical precision.

\begin{figure} 
\begin{minipage}{8in}
\epsfig{file=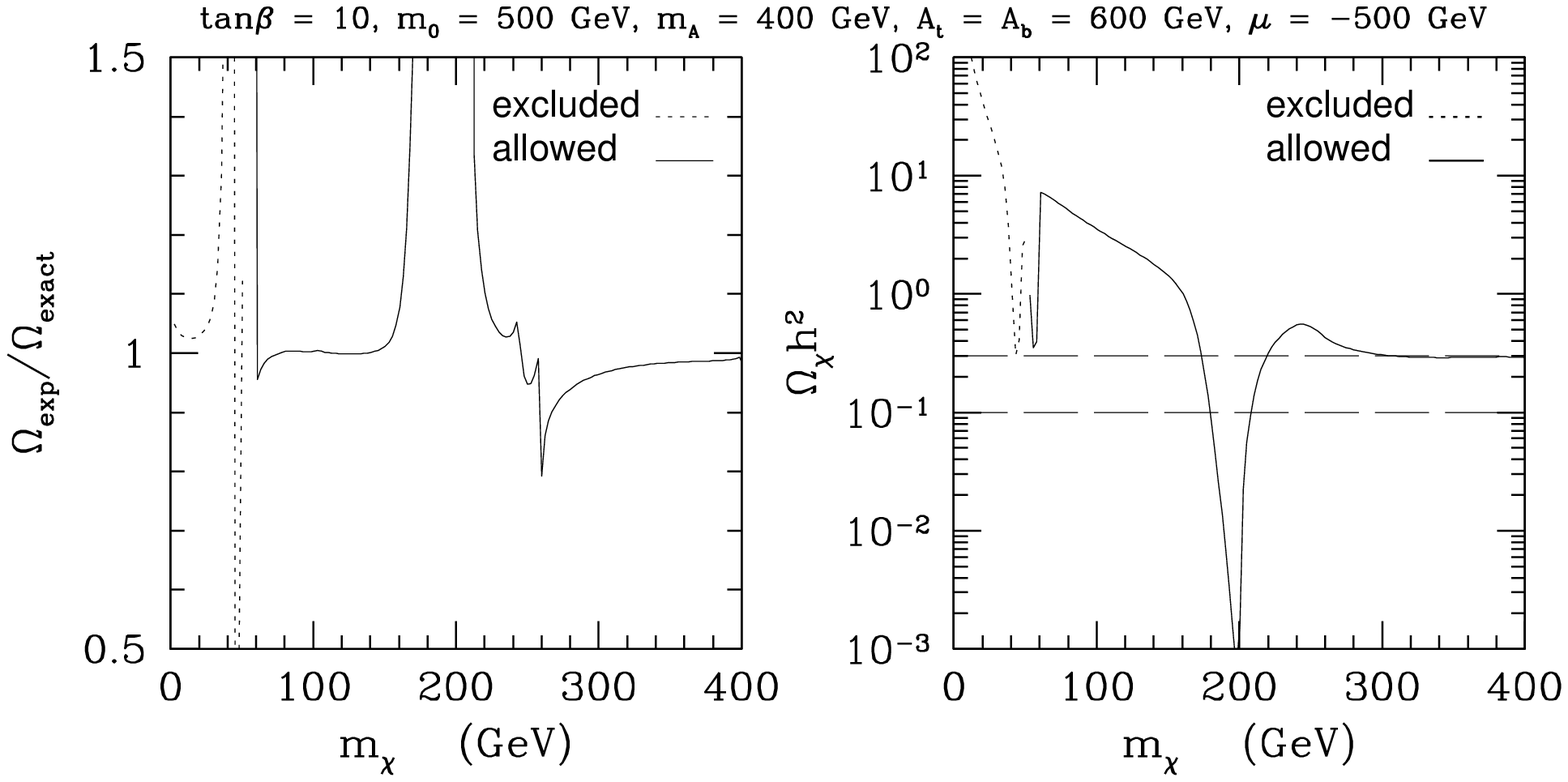,height= 6in}
\vspace*{-7.5cm}
\end{minipage}
\caption[fig2]{ 
The ratio $\Omega_{\rm exp}/\Omega_{\rm exact}$
(a) and the relic density $\Omega_{\chi} h^2$ (b) for the same
choice of parameters as in Fig.~\protect{\ref{fig:nrr-fig1}}. The
solid (dotted) curves are allowed (excluded) by current experimental
constraints. 
In
window (a) the relic aboundance in both cases is computed by solving 
Eq.~(\protect{\ref{eqn:freeze-out-temperature}}) iteratively and using
Eq.~(\protect{\ref{eqn:relic-density}}). In window~(b) we show $\Omega_{\chi}
h^2$ is computed using our numerical code. The band between the two
horizontal dashed lines corresponds to the cosmologically favoured
range $0.1<\abundchi<0.3$.
} 
\label{fig:omega10-}
\end{figure}

In Fig.~\ref{fig:omega10-}a, we plot the ratio $\Omega_{\rm
  exp}/\Omega_{\rm exact}$ versus $m_\chi$ for the same set of
parameters as before. We denote by $\Omega_{\rm
  exact}$ and $\Omega_{\rm exp}$ the neutralino 
relic density calculated with the
exact~(\ref{eqn:thermal-average}) and approximate~(\ref{expansion:eq})
formulae for the thermal average, respectively. The relic abundance
in both cases is computed by solving
Eq.~(\ref{eqn:freeze-out-temperature}) iteratively and using
Eq.~(\ref{eqn:relic-density}).
From now on we impose all applicable experimental constraints,
including the lightest Higgs mass constraint $m_h >
113.5\gev$~\cite{Higgs_mass_constraint} 
and a lower bound $104\gev$ for chargino masses.  In the case of
$b\rightarrow s \gamma$ we have followed DarkSusy in using a somewhat
old range $1.0\times 10^{-4}$ $<$ $B(b \rightarrow s \gamma)$ $<$
$4.0\times 10^{-4}$~\cite{ref:bsgold} and have not yet included 
new SUSY contributions at large $\tan\beta$ that have been recently
pointed out~\cite{bsgtanbeta:ref}. We feel that this
is sufficient for the purpose of this analysis.  The solid line
represents the experimentally allowed region while the dotted line
corresponds to the low $\mchi$ range excluded by collider experiments.

It is evident that the ratio $\Omega_{\rm exp}/\Omega_{\rm exact}$
varies considerably with $\mchi$. Near resonances it is very large as
discussed above. 
If we conservatively demand a $10\%$ accuracy in the
relic density computations then the approximate method fails to satisfy
this criterion over a broad range $160\gev\lsim\mchi\lsim220\gev$,
which is a significant fraction of the neutralino mass range. (In
fact, various naturalness constraints, even if only indicative, give a
rough constraint $\mchi\lsim 200\gev$~\cite{chiasdm,dg}.) For
larger values of $\tan\beta$ the range of $\mchi$ where $0.9<\Omega_{\rm
exp}/\Omega_{\rm exact}<1.1$ is not satisfied becomes even larger.

It is also clear that even further away from resonances the ratio
$\Omega_{\rm exp}/\Omega_{\rm exact}$ exhibits a rather complex
behavior. For $220\gev\lsim\mchi\lsim260\gev$ the new channels $W^\pm
H^\mp$, $ZA$ and $ZH^0$ successively kick in, and become actually even
somewhat more important than $f\bar{f}$. Since, as we mentioned above,
the expansion method does not work properly near new channels
thresholds~\cite{gs91}, the ratio varies quite rapidly over the
mentioned range of $\mchi$, although not as much as near resonances.
Finally, far away from resonances and thresholds, the expansion method
seems to work remarkably well.

In Fig.~\ref{fig:omega10-}b we show the relic abundance $\Omega_\chi
h^2$ as a function of $\mchi$. Again,
the solid (dotted) line represents the range of $\mchi $ allowed
(excluded) by the experimental constraints. The two horizontal dashed
lines delimit the cosmologically expected range~(\ref{eqn:omegah^2})
of $\Omega_\chi h^2$.  Two features should be mentioned. Over the
range of smaller $\mchi$, in particular near resonances, the relic
abundance varies rapidly. On the other hand, at larger $\mchi$, where
new channels involving the ``heavy'' Higgs bosons $H^\pm$, $A$ and
$H$ open up and can become important, $\Omega_{\chi} h^2$ exhibits a
wide plateau. In both regions it is clearly important to compute
$\Omega_{\chi} h^2$ accurately if one wants to achieve an accuracy
expected from future determinations of $\Omega_{\rm CDM}h^2$. In
particular, it is clear from Fig.~\ref{fig:omega10-}b that at larger
$\mchi$ a relatively small difference of a few per cent in determining
$\Omega_{\chi} h^2$ could exclude on cosmological grounds the
range of $\mchi\gsim 280\gev$ which in Fig.~\ref{fig:omega10-}b is (barely)
allowed. This again illustrates the uncertainties involved in
computing $\abundchi$ and in determining the cosmologically allowed
regions of SUSY parameters.

%
%
\section{Conclusions}\label{conclusions:sec}

While we are certainly still some time away from a ``precision era''
for measuring cosmological parameters, much effort has been focussed
on improving theoretical calculations of the relic abundance of the
neutralino. We have derived a full set of analytic expressions for the
neutralino annihilation cross sections into all tree-level final
states.  In this paper, we have presented one example of such an
expression for the neutralino annihilation into $W^+ H^-$ which can be
even dominant. We have also computed the first two terms in the expansion
of the exact formulae in powers of $x$ and we
have compared the integral of the thermally averaged product of the
cross section and velocity and the relic abundance computed both ways.
We have confirmed the well-known inaccuracy of the approximate method
near reasonances and threshold and showed that, far away from such
cases the expansion method works well. However, due to the presence of
several resonances and thresholds in the MSSM, the expansion method
may lead to sizeable errors over a relatively large range of
the neutralino mass of up to several tens of $\gev$. 
We have also demonstrated that the iterative way of
computing the freeze-out point works rather well for both the exact
and expansion methods of computing the thermal average but not in the
latter case near resonances and thresholds. 

%
%
\section*{Acknowledgements}
T.N. would like to 
thank the CERN Theory Division for its kind hospitality.

%
\section*{Appendix}
Here we give expressions for the auxiliary functions used in the text. 
First, we define
\begin{eqnarray}
   D(s,x,y_1,y_2)&\equiv& x+\frac{y_{1}+y_{2}}{2}-\frac{s}{2} \, ,\\
   F(s,x,y_1,y_2)&\equiv& \frac{1}{2}\sqrt{s-4\,x} 
\sqrt{s-(\sqrt{y_1}+\!\sqrt{y_2})^2}
\,\sqrt{1-\frac{(\sqrt{y_1}-\!\sqrt{y_2})^2}{s}} \, .
\end{eqnarray}
If we define $D\equiv D(s,x,y_{1},y_{2})$, $F\equiv F(s,x,y_{1},y_{2})$, 
$t_{\pm}(s,x,y_1,y_2) \equiv D \pm F$ 
 and $({\mathcal T}_{i},{\mathcal Y}_{i})$ $\equiv$ 
$({\mathcal T}_{i},{\mathcal Y}_{i}) (s,x,y_{1},y_{2},z_{1},z_{2})$ 
($i$ $=$ $0,1,2$), then 
\begin{eqnarray}
  {\mathcal F}(s,x,y_1,y_2,z) &=& \frac{1}{2\,F}\,
\ln\left|\frac{t_{+}(s,x,y_1,y_2)-z}{t_{-}(s,x,y_1,y_2)-z}\right|\, ,
\end{eqnarray}
and
\begin{eqnarray}
\!\!\!\!\!\!\!\!\! {\mathcal T}_{0} \! &=& \!
\frac{1}{z_1-z_2}[{\mathcal F}(s,x,y_1,y_2,z_1)
                      -{\mathcal F}(s,x,y_1,y_2,z_2)], \nonumber \\
\!\!\!\!\!\!\!\!\! {\mathcal T}_{1} \! &=& \!
\frac{1}{z_1-z_2}[z_1\,{\mathcal F}(s,x,y_1,y_2,z_1)
                     -z_2\,{\mathcal F}(s,x,y_1,y_2,z_2)], \nonumber \\
\!\!\!\!\!\!\!\!\! {\mathcal T}_{2} \! &=& \!
1+\frac{1}{z_1-z_2}[z_1^2\,{\mathcal F}(s,x,y_1,y_2,z_1)
                     -z_2^2\,{\mathcal F}(s,x,y_1,y_2,z_2)], \nonumber \\
\!\!\!\!\!\!\!\!\! {\mathcal Y}_{0} \! &=& \!
\frac{1}{z_1+z_2-2\,D}[{\mathcal F}(s,x,y_1,y_2,z_1)
                        +{\mathcal F}(s,x,y_1,y_2,z_2)], \nonumber \\
\!\!\!\!\!\!\!\!\! {\mathcal Y}_{1} \! &=& \!
\frac{1}{z_1+z_2-2\,D}[2\,(z_1-D)\,{\mathcal F}(s,x,y_1,y_2,z_1)
                   -2\,(z_2-D)\,{\mathcal F}(s,x,y_1,y_2,z_2)], \nonumber \\
\!\!\!\!\!\!\!\!\! {\mathcal Y}_{2} \! &=& \!
1 +\frac{1}{z_1+z_2-2\,D} 
[z_1\,(z_1-2\,D){\mathcal F}(s,x,y_1,y_2,z_1) \nonumber \\
& & \hspace{5cm} + \ z_2\,(z_2-2\,D)\,{\mathcal
  F}(s,x,y_1,y_2,z_2)]. \nonumber  
\end{eqnarray}
The  expressions for $G_{WH}^{T(i)}$ and $G_{WH}^{Y(j)}$, where $i=1,2,3$
and $j=1,\cdots,5$,
in Eqs.~(\ref{eqn:I_2}-\ref{eqn:I_0}) are given by
   \begin{eqnarray}
 \!\! G_{WH}^{T(1)} \! &\equiv& \! 
s (m_{\chi}^2+2m_{W}^2) - m_{\chi}^4
- m_{\chi}^2 (m_{H^{\pm}}^2+3m_{W}^2) - m_{W}^2m_{H^{\pm}}^2, \nonumber \\
 \!\! G_{WH}^{T(2)} \! &\equiv& \! s (m_{\chi}^2+2m_W^2)-2m_{\chi}^4 
+m_{\chi}^2 (m_{H^{\pm}}^2-m_W^2)-2m_W^2 (m_{H^{\pm}}^2+m_W^2), \nonumber \\
 \!\! G_{WH}^{T(3)} \! &\equiv& \! 
(m_{\chi}^2-m_{H^{\pm}}^2)\,(m_{\chi}^2-m_{W}^2)
\,(m_{\chi}^2+2\,m_{W}^2), \nonumber \\
 \!\! G_{WH}^{Y(1)} \! &\equiv& \! G_{WH}^{T(1)} \, , \nonumber \\
 \!\! G_{WH}^{Y(2)} \! &\equiv& \! 
3\,s\,m_W^2 - 12\,m_W^2 m_{\chi}^2
 + 3\,m_W^2 m_{H^{\pm}}^2 - 3\,m_W^4, \nonumber \\
 \!\! G_{WH}^{Y(3)} \! &\equiv& \! 
s^2 - s(2m_{\chi}^2+2m_{H^{\pm}}^2+3m_W^2)
 + 2m_{\chi}^4 + (2m_{\chi}^2+m_{H^{\pm}}^2)(m_{H^{\pm}}^2+3m_W^2)
 - 2m_W^4, \nonumber \\
 \!\! G_{WH}^{Y(4)} \! &\equiv& \! 
s \,(m_{\chi}^2-2 m_W^2)(m_{\chi}^2-m_{H^{\pm}}^2)
+ 4 m_W^2 m_{\chi}^4   \nonumber \\
\! & & \! + \ m_{\chi}^2 (m_{H^{\pm}}^4 - 5 m_W^2 m_{H^{\pm}}^2 + 2 m_W^4 )
- 2 m_W^4 m_{H^{\pm}}^2,  \nonumber \\
 \!\! G_{WH}^{Y(5)} \! &\equiv& \! 
s \,m_{\chi}^2 (m_{\chi}^2-m_W^2) - m_{\chi}^6 
- m_{\chi}^4 (m_{H^{\pm}}^2 + 3 m_W^2)
+ m_{\chi}^2 m_W^2 (2 m_{H^{\pm}}^2 + 3 m_W^2).\nonumber 
\end{eqnarray}

\vspace*{1cm}

%
%
%
%
\end{document}